\documentclass[a4paper,11pt]{article}
\usepackage{pos}
\title{Generating configurations of increasing lattice size with machine learning and the inverse renormalization group}

\author*[a]{Dimitrios Bachtis}

\affiliation[a]{Laboratoire de Physique de l'Ecole Normale Sup\'erieure, ENS, Universit\'e PSL,
CNRS, Sorbonne Universit\'e, Universit\'e de Paris, F-75005 Paris, France}
\emailAdd{dimitrios.bachtis@phys.ens.fr}
\abstract{We review recent developments of machine learning algorithms pertinent to the inverse renormalization group, which was originally established as a generative numerical method by Ron-Swendsen-Brandt via the implementation of compatible Monte Carlo simulations. Inverse renormalization group methods enable the iterative generation of configurations for increasing lattice size without the critical slowing down effect. We discuss the construction of inverse renormalization group transformations with the use of convolutional neural networks and present applications in models of statistical mechanics, lattice field theory, and disordered systems. We highlight the case of the three-dimensional Edwards-Anderson spin glass, where the inverse renormalization group can be employed to construct configurations for lattice volumes that have not yet been accessed by dedicated supercomputers. }

\FullConference{European network for Particle physics, Lattice field theory and Extreme computing (EuroPLEx2023)\\
 11-15 September 2023\\
Berlin, Germany\\}

\begin{document}

\include{ms.bib}

\maketitle

\section{Inverse Monte Carlo renormalization group}

Inverse renormalization group methods were originally established as generative numerical techniques by Ron-Swendsen-Brandt via the implementation of compatible Monte Carlo simulations~\citep{PhysRevLett.89.275701}. The authors implemented the method on configurations of the three-dimensional Ising model with lattice volume $V=4^{3}$  to construct inversely renormalized systems up to lattices of $V=128^{3}$. The discussed approach relies on the implementation of a Monte Carlo technique to ensure a computationally stable inverse transformation, is shown to evade the critical slowing down effect for the case of the three-dimensional Ising model, and provides accurate calculations of the critical exponents. 

Machine learning algorithms have recently revolutionized multiple aspects of academia and industry, and a natural question is whether neural networks could be employed to confront the mathematically ill-defined problem of constructing inverse renormalization group transformations. A straightforward approach would then consider the application of a standard real-space transformation with a rescaling factor of $b$, such as the majority rule, on original configurations of a given lattice size $L$ to construct renormalized configurations of lattice size $L'=L/b$ . One could then present the renormalized configurations of $L'$ as input to a machine learning algorithm, such as a set of (transposed) convolutions, in order to approximately reproduce the original configurations of size $L$. 

This implementation then defines the construction of a kernel for an inverse transformation which can be arbitrarily optimized to approximate the inversion of a standard renormalization group transformation, such as the majority rule, by increasing the set of variational parameters, namely the weights and biases of the machine learning algorithm. The benefit of the discussed approach is that if one considers only a set of convolutions in the machine learning implementation, which can be applied irrespective of the lattice size, one could then increase the volume of the system for an arbitrary number of times via consecutive applications of the inverse transformation.

\begin{figure}[b]
\centering
\includegraphics[width=6cm]{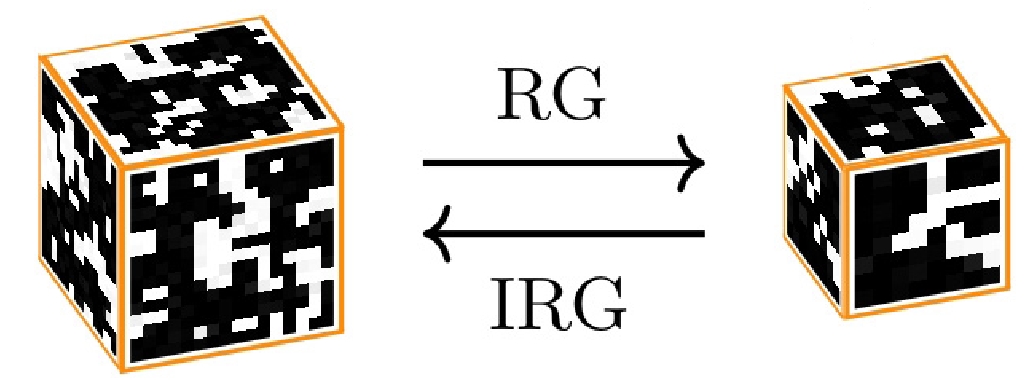}
\caption{\label{fig:1} Illustration of the application of a standard renormalization group transformation (RG) with the majority rule and its approximate inversion (IRG) with the implementation of machine learning algorithms. }
\end{figure}

Inverse renormalization group transformations have been constructed with a variety of machine learning architectures, including convolutional neural networks, in Ising and Potts models~\citep{PhysRevB.99.075113,Shiina2021,PhysRevLett.121.260601}. Alternative constructions of inverse transformations are also obtained via the wavelet-conditional-renormalization group~\citep{PhysRevX.13.041038}. Conceptually related approaches, which do not consider explicit constructions of inverse transformations, but combine insights from the renormalization group and multigrid methods, have been  applied in lattice gauge theories~\citep{PhysRevD.92.114516}.

In this manuscript, we discuss the approximate construction of inverse renormalization group transformations for systems with continuous degrees of freedom, a result first demonstrated in the  context of the $\phi^{4}$ lattice field theory~\citep{PhysRevLett.128.081603}. We then present recent results on the construction of inverse renormalization group transformations for disordered systems~\citep{bachtis2023inverse}, which enable the iterative generation of configurations for lattice volumes that have not yet been accessed by dedicated supercomputers.

\section{Inverse renormalization group of quantum field theories}

We consider the two-dimensional $\phi^{4}$ scalar field theory on a square lattice:
\begin{equation}
S=- \kappa_{L} \sum_{\langle ij \rangle} \phi_{i} \phi_{j} + \frac{(\mu_{L}^{2}+4 \kappa_{L})}{2} \sum_{i} \phi_{i}^{2} + \frac{\lambda_{L}}{4} \sum_{i} \phi_{i}^{4},
\end{equation}
where $\kappa_{L}$, $\mu_{L}^{2}$, $\lambda_{L}$ are dimensionless parameters. For positive and fixed values of $\lambda_{L}>0$ and $\kappa_{L}>0$, the system transitions from a symmetric to a broken-symmetry phase for a unique negative value of $\mu^{2}_{L} < 0$. The $\phi^{4}$ lattice field theory is simulated based on the Brower-Tamayo approach~\citep{PhysRevLett.62.1087}, specifically by considering a combination of the Metropolis and Wolff algorithms. Here we consider $\lambda_{L}=0.7$, $\kappa_{L}=1$ and denote $\mu^{2}_{L} \equiv K$, where $K_{c}$ is the critical coupling.

To construct an inverse transformation we must first devise a standard real-space transformation for the $\phi^{4}$ theory. We simulate the system for $\mu_{L}^{2} = -0.9515$ to obtain a set of original configurations for lattice size $L=32$ in each dimension. We separate the lattice into blocks of size $b\times b$, where $b=2$ is the rescaling factor and we sum the degrees of freedom within each block. If the sum is positive (negative)  we select the renormalized degree of freedom as being equal to the mean of the positive (negative) degrees of freedom. An important observation is that by reducing the lattice size $L$ by a rescaling factor of $b$ as $L'=L/b$ within the context of the standard renormalization group, we are simultaneously reducing the correlation length $\xi$ by an equal factor, namely $\xi'=\xi/b$. Once we have approximated the inversion of the standard renormalization group transformation we can apply it, in principle, for an arbitrary number of times to iteratively increase the lattice size $L$ of the system. Thus in the context of the inverse renormalization group, the inversely renormalized lattice size is $L'=bL$ with $\xi'=b\xi$. 

By observing how the correlation length is transformed in the standard and in the inverse renormalization group we are able to obtain an intuitive understanding of the induced standard and inverse renormalization group flows. Let us consider the case of the two-dimensional $\phi^{4}$ theory. In the case of the standard renormalization group the application of each transformation reduces the correlation length and as a result the renormalized system is driven deeper into either the symmetric or the broken-symmetry phase. Conversely, in the case of the inverse renormalization group the application of each inverse transformation increases the correlation length of the system, thus driving it closer to the fixed point. These observations are established by considering a one-dimensional renormalization group flow or, equivalently, under the consideration that the renormalized system is an accurate representation of the original at a different value of the $\mu^{2}$ coupling. An important insight is then that the correlation length diverges at the critical coupling $\mu_{c}^{2}$ and it is exactly this observation that provides a self-consistent approach to locate the critical fixed point in Monte Carlo renormalization group methods: it is the point in parameter space where intensive observables intersect.

\begin{figure}[t]
\centering
\includegraphics[width=9cm]{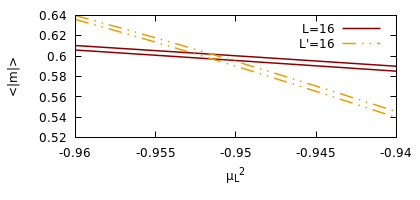}
\caption{\label{fig:2}  Absolute value of the magnetization versus the squared mass for an original $L$ and a renormalized $L'$ system of identical lattice size. The renormalized system has been obtained via the application of a standard renormalization group transformation. Figure from Ref.~\citep{PhysRevLett.128.081603}.}
\end{figure}

An application of the standard renormalization group in the case of the $\phi^{4}$ theory is provided in Fig.~\ref{fig:2}, which depicts the absolute value of the intensive magnetization for an original and a renormalized system of identical lattice size $L=16$. We observe that all of the qualitative behaviour discussed in the previous paragraph is now reproduced. Specifically for $\mu^{2}<\mu_{c}^{2}$, $|m'|>|m|$ and equivalently $\mu^{2}>\mu_{c}^{2}$, $|m'|<|m|$. In addition, we observe an intersection for the intensive magnetization which provides an estimate of the critical fixed point $\mu_{c}^{2}$.

\begin{figure}[t]
\centering
\includegraphics[width=12cm]{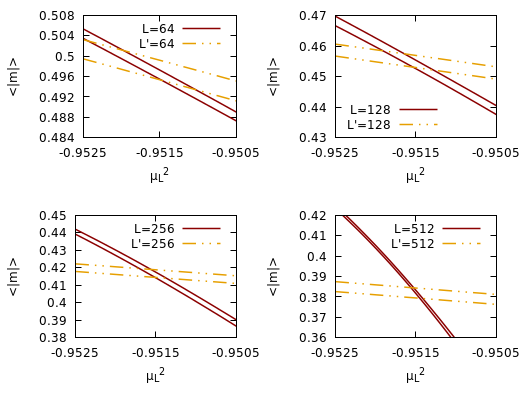}
\caption{\label{fig:3}  Absolute value of the magnetization versus the squared mass for an original $L$ and a renormalized $L'$ system of identical lattice size. The renormalized systems have been obtained via the application of an inverse renormalization group transformation.  Figure from Ref.~\citep{PhysRevLett.128.081603}.}
\end{figure}

In Fig.~\ref{fig:3} we depict results obtained with the inverse renormalization group. Specifically we start with a lattice size of $L=32$ and apply the inverse renormalization group iteratively until we construct configurations of lattice size $L'=512$. We then compare the inversely renormalized system against an original system of identical lattice size~\citep{arxiv.2205.08156}. We observe the qualitative behaviour discussed above for the inverse renormalization group, specifically  for $\mu^{2}>\mu_{c}^{2}$, $|m'|<|m|$ and equivalently $\mu^{2}<\mu_{c}^{2}$, $|m'|>|m|$. In addition we locate an intersection of observables in parameter space which corresponds to the critical fixed point. We then utilize all of the renormalized systems, up to $L'=512$ to extract two critical exponents. The results, which are in agreement with the two-dimensional universality class, are provided in Ref.~\citep{PhysRevLett.128.081603}.

\section{Inverse renormalization group of disordered systems}

We consider as an example the three-dimensional Edwards-Anderson spin glass with replicas $\sigma$, $\tau$ which comprise spins $s$, $t$. The two-replica Hamiltonian is given by:
\begin{equation}\label{eq:origham}
E_{\sigma,\tau} = E_{\sigma}+E_{\tau}=-\sum_{\langle ij \rangle} J_{ij} (s_{i}s_{j}+t_{i}t_{j}),
\end{equation}
where $s,t=\pm 1$, $\langle ij \rangle$ are nearest-neighbors $i$ and $j$, $J_{ij}$ is randomly sampled as $J_{ij}=\pm1$, and $\lbrace J_{ij} \rbrace$ is a realization of disorder.

\begin{figure}[t]
\centering
\includegraphics[width=9cm]{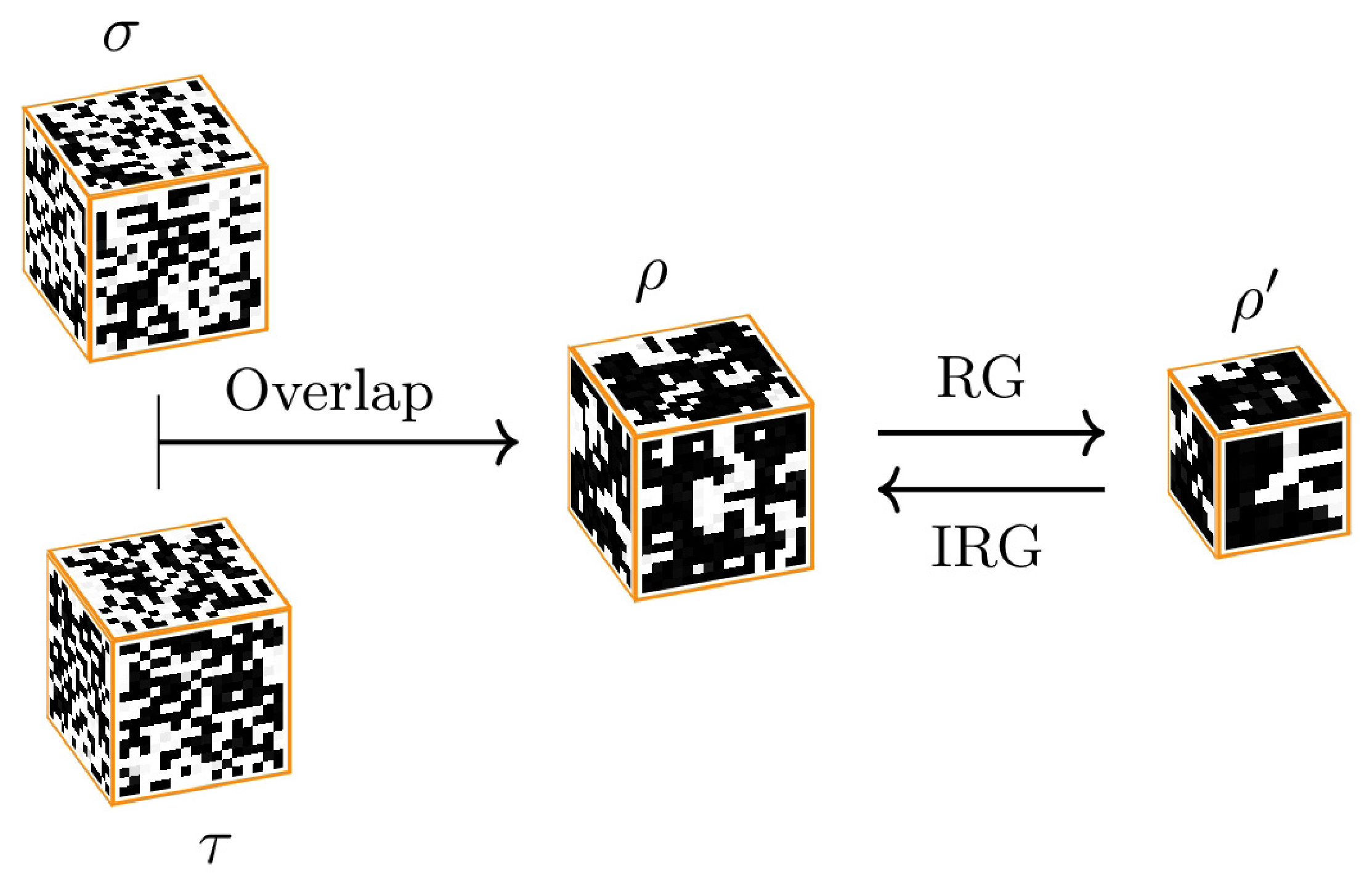}
\caption{\label{fig:4} The transition to the overlap configurations for the case of the three-dimensional Edwards-Anderson model and the implementation of a standard renormalization group transformation on the effective spin glass which comprises overlap degrees of freedom. Figure from Ref~\citep{bachtis2023inverse}. }
\end{figure}

The spin glass phase transition of the system can be studied by the overlap order parameter which is defined over two replicas $\sigma$, $\tau$:
\begin{equation}
q_{\sigma\tau}= \frac{1}{V} \sum_{i} s_{i} t_{i},
\end{equation} 
where $V=L^{3}$ is the volume of the system, $L$ the lattice size in each dimension, and $\rho_{i}=s_{i} t_{i}$ defines an overlap variable.

The concept of the renormalization group is nontrivial for the case of disordered systems. Since we are interested in studying the spin glass phase transition based on the overlap order parameter, we must devise a renormalization group transformation that is capable of properly transforming the overlap order parameter. This implies that we require an effective Hamiltonian that is exclusively a function of the overlap variables $\rho$. This type of effective Hamiltonian, implemented to describe the effective spin glass, was first introduced by Haake-Lewenstein-Wilkens~\citep{PhysRevLett.55.2606} and defines an effective probability distribution
\begin{equation}\label{eq:HLW}
P_{\rho_{i}} =   2^{V} \Bigg[  \frac{\exp\big[\beta \sum_{\langle ij \rangle} J_{ij}(1+\varrho_{i}\varrho_{j})\big]}{Z^{2}[\lbrace J_{ij} \rbrace]} \Bigg]_{J_{ij}}.
\end{equation}
We remark that $\langle \rangle$ defines a thermal average and $[]$ is an averaging over the realizations of disorder. The Haake-Lewenstein-Wilkens approach provides a formal mathematical mapping which maps the spin glass phase transition of the original Hamiltonian into a phase transition that resembles ferromagnetic ordering in the effective Hamiltonian of the overlap system. Renormalization group transformations on this effective Hamiltonian~\citep{PhysRevB.37.7745,2302.08459} were first established for the effective spin glass by Wang and Swendsen~\citep{PhysRevB.37.7745}. Inverse renormalization group transformations  were  introduced in Ref.~\citep{bachtis2023inverse}. The concept of transitioning to the overlap configurations, applying and inverting the renormalization group is illustrated in Fig.~\ref{fig:4}. 

\begin{figure}[t]
\centering
\includegraphics[width=15cm]{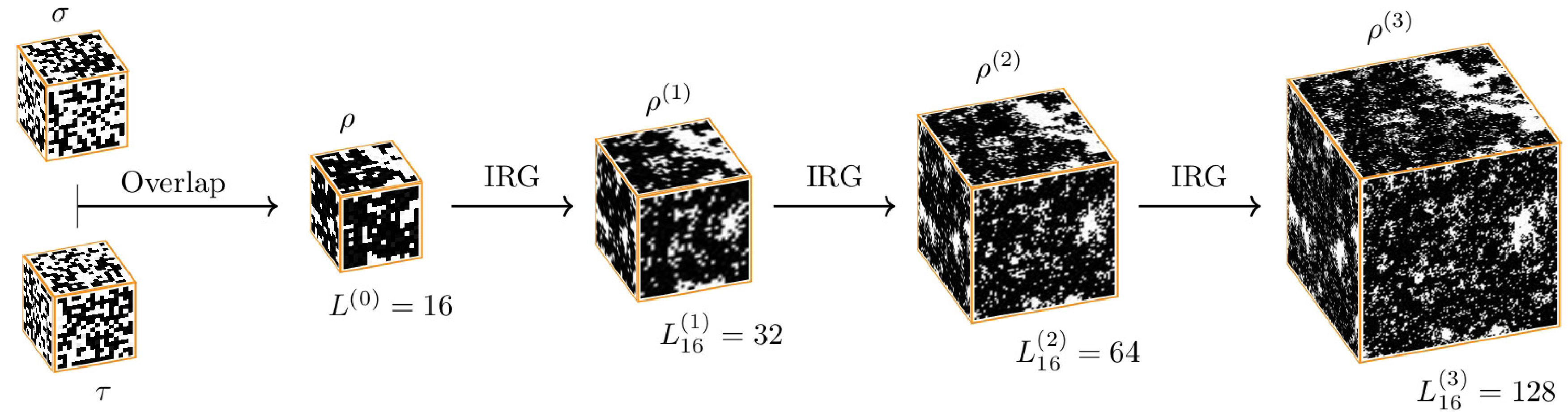}
\caption{\label{fig:5} The iterative application of inverse renormalization group transformations for the case of the three-dimensional Edwards-Anderson. Starting from a lattice size of $L=16$ we apply the inverse transformations until we construct $L'=128$. Figure from Ref~\citep{bachtis2023inverse}. }
\end{figure}

We remark that spin glasses have posed some of the greatest challenges to the community of computational physics in relation to how numerically challenging these systems are to simulate. These difficulties stem from the necessity of implementing replica exchange Monte Carlo methods~\citep{PhysRevLett.57.2607} on a wide range of temperatures, from the simulation of multiple real replicas, and from the need to extensively monitor the equilibration of Markov chains.  To provide direct evidence, one needs only observe that the largest lattice volume simulated for the three-dimensional Edwards-Anderson model is $V=40^{3}$, and this has been achieved exclusively via the use of special-purpose machines, namely the dedicated Janus supercomputer~\citep{PhysRevB.88.224416}. Consequently, it is natural to ask whether one can implement the inverse renormalization group, which enables the instant generation of equilibrated configurations for increasing lattice size, to construct configurations for lattice volumes that have not yet been accessed by supercomputers.

An illustration of the inverse renormalization group applied iteratively in the case of spin glasses to arbitrarily increase the lattice size is provided in Fig.~\ref{fig:5}. Starting from lattices of size $L=16$ we apply the inverse transformation to construct lattices of size $L=128$. The inversely renormalized configurations can then be utilized to extract critical exponents of the system, and the results are provided in Ref.~\citep{bachtis2023inverse}.

\section{Conclusions}

We have briefly reviewed inverse renormalization group methods, which provide a computationally efficient construction of equilibrated configurations for increasing lattice size in absence of the critical slowing down effect. Specifically, we presented applications of the inverse renormalization group in the research fields of lattice field theory and of disordered systems.

 Monte Carlo renormalization group methods~\citep{PhysRevLett.42.859} provide well-defined standards to systematically improve calculations of critical exponents and to reduce pertinent errors. We remark that it is conceptually straightforward to incorporate numerical exactness within inverse renormalization group methods in order to enable the construction of exact configurations for increasing lattice volumes that have not yet been accessed by dedicated supercomputers. This direction might be particularly appealing in the context of disordered systems, given the prohibitively long simulations required to sample these systems. Besides providing potential improvements in the sampling of statistical systems, inverse renormalization group methods could potentially be extended to new venues, for instance within the study of the phase transitions which emerge during the training of machine learning algorithms~\citep{bachtis2024cascade}.

\section{Acknowledgements}

The author has received funding from the European Research Council (ERC) under the European Union’s Horizon 2020 research and innovation programme under grant agreement No 813942. The author acknowledges support from the CFM-ENS Data Science Chair and PRAIRIE (the PaRis Artificial Intelligence Research InstitutE).

\bibliography{ms}

%

\end{document}